\RequirePackage{ifpdf}
\ifpdf 
\documentclass[pdftex]{sigma}
\else
\documentclass{sigma}
\fi

\numberwithin{equation}{section}

\def \CHL {\mathrm{CHL}}
\def \I {\mathcal{I}}
\def \cj{{\cal J}}
\def\D{\Delta}
\def\d{\delta}
\def\L{\Lambda}
\def\l{\lambda}
\def\G{\Gamma}
\def\g{\gamma}
\def\e{\epsilon}
\def\s{\sigma}

\def\O{\Omega}
\def\a{\alpha}
\def\b{\beta}

\def\s{\sigma}

\def\f{\phi}

\def\e{\epsilon}

\newcommand{\be}{\begin{equation}}
\newcommand{\ee}{\end{equation}}
\newcommand{\bea}{\begin{eqnarray}}
\newcommand{\eaa}{\end{eqnarray}}
\newtheorem{Restriction}{Restriction}
\newtheorem{Theorem}{Theorem}

\begin{document}

\allowdisplaybreaks

\renewcommand{\PaperNumber}{094}

\FirstPageHeading

\ShortArticleName{Four-Dimensional Spin Foam Perturbation Theory}

\ArticleName{Four-Dimensional Spin Foam Perturbation Theory}

\Author{Jo\~{a}o FARIA MARTINS~$^\dag$ and Aleksandar MIKOVI\'{C}~$^{\ddag\S}$}  

\AuthorNameForHeading{J. Faria Martins and A. Mikovi\'{c}}

\Address{$^\dag$~Faculdade de Ci\^{e}ncias e Tecnologia,   Universidade Nova de Lisboa, \\
\hphantom{$^\dag$}~Quinta da Torre, 2829-516 Caparica, Portugal}
\EmailD{\href{mailto:jn.martins@fct.unl.pt}{jn.martins@fct.unl.pt}}
\URLaddressD{\url{http://ferrari.dmat.fct.unl.pt/personal/jnm/}}

\Address{$^\ddag$~Departamento de Matem\'atica, Universidade Lus\'{o}fona de Humanidades e Tecnologia,\\
\hphantom{$^\ddag$}~Av do Campo Grande, 376, 1749-024 Lisboa, Portugal}
\EmailD{\href{mailto:amikovic@ulusofona.pt}{amikovic@ulusofona.pt}}

\Address{$^\S$~Grupo de F\'{\i}sica Matem\'atica da Universidade de Lisboa,\\
\hphantom{$^\S$}~Av. Prof. Gama Pinto, 2, 1649-003 Lisboa,
Portugal}

\ArticleDates{Received June 03, 2011, in f\/inal form September 23, 2011;  Published online October 11, 2011}

\Abstract{We def\/ine a four-dimensional spin-foam perturbation theory for the ${\rm BF}$-theory with a $B\wedge B$ potential term def\/ined for a compact semi-simple Lie group $G$ on a compact orientable 4-manifold $M$. This is done by using the formal spin foam perturbative series coming from the spin-foam generating functional. We then regularize the terms in the perturbative series by passing to the category of representations of the quantum group $U_q (\mathfrak{g})$ where $\mathfrak{g}$ is the Lie algebra of $G$ and $q$ is a root of unity.
The Chain--Mail formalism can be used to calculate the perturbative terms when the vector space of intertwiners $\L\otimes \L \to A$, where $A$ is the adjoint representation of $\mathfrak{g}$, is 1-dimensional for each irrep $\L$.  We calculate the partition function $Z$ in the dilute-gas limit for a special class of triangulations of restricted local complexity, which  we conjecture to exist on any 4-manifold~$M$. We prove that  the f\/irst-order perturbative contribution vanishes for f\/inite triangulations, so that we def\/ine a dilute-gas limit by using the second-order contribution. We show that $Z$ is an analytic continuation of the Crane--Yetter partition function. Furthermore, we relate $Z$ to the partition function for the $F\wedge F$ theory.}


\Classification{81T25; 81T45; 57R56}

\section{Introduction}

Spin foam models are state-sum representations of the path integrals for ${\rm BF}$ theories on  simplicial complexes. Spin foam models are used to def\/ine topological quantum f\/ield theories and quantum gravity theories, see \cite{Ba}. However, there are also perturbed ${\rm BF}$ theories in various dimensions, whose potential terms are powers of the $B$ f\/ield, see \cite{FK}. The corresponding spin-foam perturbation theory generating functional was formulated in \cite{FK}, {but further} progress was hindered by the lack of the regularization procedure for the corresponding perturbative expansion and the problem of implementation of the triangulation independence.

The problem of implementation of the triangulation independence for general spin foam perturbation theory was studied in \cite{Ba2}, and a solution was proposed, in the form of calculating the perturbation series in a special limit. This limit was called the dilute-gas limit, and it was given by $\l\to 0$, $N\to\infty$, such that $g=\l N$ is a f\/ixed constant, where $\l$ is the perturbation theory parameter, also called the coupling constant, $N$ is the number of $d$-simplices in a simplical decomposition of a $d$-dimensional compact manifold $M$ and $g$ is the ef\/fective perturbation parameter, also called the renormalized coupling constant. However, the dilute-gas limit could be used in a concrete example only if one knew how to regularize the perturbative contributions.

The regularization problem has been solved recently in the case of three-dimensional (3d) Euclidean quantum gravity with a cosmological constant \cite{MM2}, following the approach of \cite{BGM,MM}. The 3d Euclidean classical gravity theory is  equivalent to the ${\rm SU}(2)$ ${\rm BF}$-theory with a $B^3$ perturbation, and the corresponding spin foam perturbation expansion can be written by using the Ponzano--Regge model. The terms in this series can be regularized by replacing all the spin-network evaluations with the corresponding quantum spin-network evaluations {at a root of unity}. By using the Chain--Mail formalism \cite{Ro} one can calculate the quantum group perturbative corrections, and show that the f\/irst-order correction vanishes \cite{MM2}. Consequently, the dilute-gas limit has to be modif\/ied so that $g=\l^2 N $ is the ef\/fective perturbation parameter \cite{MM2}.

Another result of \cite{MM2} was to show that the dilute gas limit cannot be def\/ined for an arbitrary class of triangulations of the manifold. One needs a restricted class of triangulations such that the number of possible isotopy classes of a graph def\/ined from the perturbative insertions is bounded. In 3d this can be achieved by using the triangulations coming from the barycentric subdivisions of a regular cubulation of the manifold \cite{MM2}.

In this paper we are going to def\/ine the four-dimensional (4d) spin-foam perturbation theory by using the same approach and the techniques as in the 3d case. We start from a BF-theory with a $B\wedge B$ potential term def\/ined for a compact semi-simple Lie group $G$ on a compact 4-manifold $M$. In Section~\ref{s2} we def\/ine the formal spin foam perturbative series by using the spin-foam generating functional method. We then regularize the terms in the series by passing to the category of representations for the quantum group $U_q (\mathfrak{g})$ where $\mathfrak{g}$ is the Lie algebra of $G$ and $q$ is a root of unity. In Sections~\ref{soc} and~\ref{five} we then use the Chain--Mail formalism to calculate the perturbative contributions. The f\/irst-order perturbative contribution vanishes, so that we def\/ine the dilute-gas limit in Section~\ref{section5} by using the second-order contribution. We calculate the partition function $Z$ in the dilute-gas limit for a class of  triangulations of a 4-dimensional manifold which are arbitrarily f\/ine and have a controllable local complexity. We conjecture that such a class of triangulations always exists for any 4-dimensional manifold, and {can be} given  by the triangulations corresponding to the  barycentric subdivisions of a f\/ixed cubulation of the manifold. We then show that $Z$ is given as an analytic continuation of the Crane--Yetter partition function.
In Section~\ref{section6} we relate the path-integral for the $F\wedge F$ theory with the spin foam partition function and in Section~\ref{section7} we present our conclusions.

\section{Spin foam perturbative expansion}\label{s2}

Let $\mathfrak{g}$ be the Lie algebra of a semisimple compact Lie group $G$. The action for a perturbed $\rm{BF}$-theory in 4d can be written as
\begin{gather}
S= \int_M \left( B^{I} \wedge F_{I}  + \l\,g_{IJ}\, B^{I} \wedge B^{J} \right), \label{abfp}
\end{gather}
where $B=B^{I}L_{I}$ is a {$\mathfrak{g}$-valued} two-form, $L_{I}$ is a basis {of} $\mathfrak{g}$, $F= dA + \frac{1}{2}[A, A]$ is the curvature 2-form for the $G$-connection $A$ on a principal $G$-bundle over $M$, $X^I = X_I$ and $g_{IJ}$ is a symmetric $G$-invariant tensor. Here if $X$ and $Y$ are vector f\/ields in the manifold $M$ then $[A, A](X,Y)=[A(X),A(Y)]$.

We will consider the case when $g_{IJ} \propto \d_{IJ}$, where $\d_{IJ}$ is the Kronecker delta symbol. In the case of a simple Lie group, this is the only possibility, while in the case of a semisimple Lie group one can also have $g_{IJ}$ which {are} not proportional to $\d_{IJ}$. For example, in the case of
the $SO(4)$ group one can {put} $g_{ab,cd}=\e_{abcd}$, where $\e$ is the totally antisymmetric tensor and $1\le a,\dots,d \le 4$. We will also use the notation ${\rm Tr}\,(X Y)= X^I Y_I$ and $\langle X Y \rangle = g_{IJ}  X^{I} Y^{J}$.

Consider the path integral
\begin{gather}
Z(\l,M) = \int {\cal D} A {\cal D} B   e^{i\int_M \left( B^{I} \wedge F_{I}  + \l  \langle B \wedge B\rangle \right)} .\label{pibf}
\end{gather}
It can be evaluated perturbatively in $\l$ by using the generating functional
\begin{gather} Z_0({\cal J},M) = \int {\cal D} A {\cal D} B   e^{i\int_M \left( B^{I} \wedge F_{I} + B^{I}\wedge {\cal J}_I \right)} ,\label{gf}\end{gather}
(where ${\cal J}_IL^I$ is an arbitrary 2-form valued in $\mathfrak{g}$)
and the formula
\begin{gather} Z(\l,M) = \exp\left( -i\l\int_M g_{IJ}\frac{\d}{ \d {\cal J}_I}\wedge \frac{\d}{ \d {\cal J}_J}\right)Z_0({\cal J},M)\Big |_{{\cal J}=0} .\label{pif}
\end{gather}

The path integrals (\ref{gf}) and (\ref{pif}) can be represented as spin foam state sums by discretizing the 4-manifold $M$, see~\cite{FK}. This is done by using a simplicial decomposition (triangulation) of~$M$,~$T(M)$. It is useful to introduce the dual cell complex $T^*(M)$ \cite{RS} (a cell decomposition of~$M$), and we will denote the vertices, edges and faces of $T^*(M)$ as $v$, $l$ and $f$, respectively. A~vertex $v$ of $T^*(M)$ is dual to a 4-simplex $\s$ of $T(M)$, an edge $l$ of $T^*(M)$ is dual to a~tetrahedron~$\tau$ of~$T(M)$ and a face~$f$ of~$T^*(M)$ is dual to a triangle~$\D$ of~$T(M)$.

The action (\ref{abfp}) then takes the following form on $T(M)$
\begin{gather*}
S= \sum_\D {\rm Tr}\left(B_\D F_f\right) + \frac{\l}{5} \sum_{\s} \sum_{\D',\D''\in \s}\langle B_{\D'} B_{\D''} \rangle,
\end{gather*}
where $\D'$ and $\D''$ are pairs of triangles in a four-simplex $\s$ whose intersection is a single vertex of $\s$ and $B_\D = \int_\D B$. The variable $F_f$ is def\/ined as
\begin{gather*} e^{F_f} = \prod_{l\in \partial f}g_l ,
\end{gather*}
where $f$ is the face dual to a triangle $\D$, $l$'s are the edges of the polygon boundary of $f$ and $g_l$ are the dual edge holonomies.

One can then show that
\begin{gather}
 Z_0 ({\cal J},M) = \sum_{\L_f, \iota_l} \prod_f \dim \L_f \prod_v A_5 (\L_{f(v)} , \iota_{l(v)} ,
{\cal J}_{f(v)}) ,\label{sfgf}\end{gather}
where the amplitude $A_5 (\L_{f(v)} , \iota_{l(v)} , {\cal J}_{f(v)})$, also called the weight for the 4-simplex~$\s$, is given by the evaluation of the four-simplex spin network whose edges are colored by ten $\L_{f(v)}$ irreps and f\/ive $\iota_{l(v)}$ intertwiners, while each edge has a $D^{(\L)}(e^{\cal J})$ insertion. Here $D^{(\L)}(e^{\cal J})$ is the representation matrix for a group element $e^{\cal J}$ in the irreducible representation (irrep) $\L$, see~\mbox{\cite{FK,M3}}. Note that a  vertex $v$ is dual to a 4-simplex $\s$, so that the set of faces $f(v)$ intersecting at~$v$ is dual to the set of ten triangles of~$\s$. Similarly, the set of f\/ive dual edges~$l(v)$ intersecting at~$v$ is dual to the set of f\/ive tetrahedrons of $\s$.
The sum in~(\ref{sfgf}) is over all colorings~$\L_f$ of the set of faces $f$ of $T^*(M)$  by the irreps $\L_f$ of $G$, as well as over the corresponding intertwiners~$\iota_l$ for the dual complex edges~$l$. Equivalently, $\L_f$ label the triangles of $T(M)$, while $\iota_l$ label the tetrahedrons of $T(M)$.

In the case of the $SU(2)$ group and ${\cal J} =0$, the amplitude $A_5$ gives the $15j$ symbol, see~\cite{Ba,KL}. For the general def\/inition  of  $15j$-symbols $A_5(\L_{f(v)} , \iota_{l(v)})$ see~\cite{Ma1,Ma2}. Then $Z_0 ({\cal J}=0,M) $ can be written as
\begin{gather}\label{CY}
 Z_0(M)=\sum_{\L_f, \iota_l} \prod_f \dim \L_f \prod_v A_5 (\L_{f(v)} , \iota_{l(v)}),
\end{gather}
which after  quantum group regularization (by passing to  a root of unity) becomes a manifold invariant known as the Crane--Yetter invariant~\cite{CY}.

The formula (\ref{pif}) is now given by the discretized expression
\begin{gather}
{Z(\l,M,T)} = \exp\left( -\frac{i\l}{5}\sum_\s \sum_{\D , \D' \in \s}g^{IJ}\frac{\partial^2}{ \partial {\cal J}_f^{I} \partial {\cal J}_{f'}^{J}}\right)Z_0 ({\cal J},M){\Big |}_{{\cal J}=0} ,\label{sfpi}
\end{gather}
where $T$ denotes the triangulation of $M$. The equation (\ref{sfpi}) can be rewritten as
\begin{gather}
{Z(\l,M,T)} = \sum_{\L_f, \iota_l}\prod_f \dim \L_f \exp\left( -i\l\sum_\s \hat V_\s \right)\prod_v A_5 (\L_{f(v)} , \iota_{l(v)}),\label{osfpi}
\end{gather}
where the operator $\hat V_\s$ is given by
\begin{gather} \hat V_\s = \frac{1}{5}\sum_{\D , \D' \in \s} g^{IJ} L^{(\L)}_{I} \otimes L^{(\L')}_{J} \equiv \frac{1}{5}\sum_{f, f';v \in f \cap f'} g^{IJ} L^{(\L_f)}_{I} \otimes L^{(\L_{f'})}_{J}.\label{vop}
\end{gather}
This operator acts on the $\s$-spin network evaluation $A_5$ by inserting the Lie algebra basis element~$L^{(\L)}$ for an irrep $\L$ into the  spin network edge carrying the irrep~$\L$. The expression (\ref{vop}) follows from (\ref{sfgf}), (\ref{sfpi}) and the relation
\[
 \frac{\partial D^{(\L)}(e^{\cal J})}{ \partial {\cal J}^I}{\Big |}_{{\cal J}=0} = L_I^{(\L)}.
\]

Following \eqref{vop}, let us def\/ine a $g$-edge in a 4-simplex spin network as a line connecting the middle points of two edges of the spin network, such that this line is labelled by the tensor $g_{IJ}$. We associate to  a $g$-edge the linear map
\[
\sum_{IJ}  g^{IJ}  L^{(\L_f)}_{I} \otimes L^{(\L_{f'})}_{J} ,
\]
where $\L_f$ and $\L_{f'}$ are the labels of the spin network edges connected by the $g$-edge and $L^{(\L_f)}_{I}$ denotes the action of the basis element $L_I$ of $\mathfrak{g}$ in the representation $\Lambda_f$.

The action of the operator $g^{IJ} L^{(\L_f)}_{I} \otimes L^{(\L_{f'})}_{J}$ in a single 4-simplex of the ${\cal J}=0$ spin foam state sum (\ref{sfgf}) can be represented as the evaluation of a spin network $\G_{\s,g}$ obtained from the 4-simplex spin network~$\G_\s$ by adding a $g$-edge between the two edges of~$\G_\s$ labeled by~$\L_f$ and~$\L_{f'}$.

\begin{figure}[t]
\centering
\includegraphics{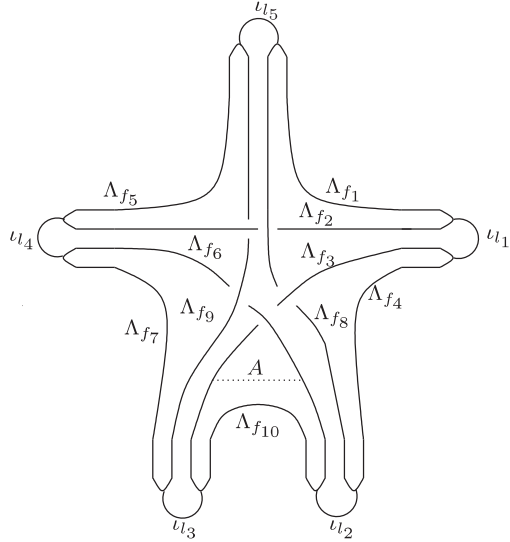}
\caption{A $15j$ symbol (4-simplex spin network) with a g-edge insertion (dashed line). Here $A$ is the adjoint representation.}
\label{net}
\end{figure}

\begin{figure}[t]
\centering
\includegraphics{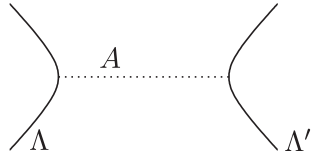}
\caption{Spin network form of equation (\ref{heval}).}
\label{insertion}
\end{figure}

When $g_{IJ} \propto \d_{IJ}$ and the intertwiners $C^{\L\L'A}$, from $\L \otimes \L'$ to $A$, where $A$ is the adjoint representation,  form a  one-dimensional vector space, $\G_{\s,g}$ becomes the 4-simplex spin network with an insertion of an edge labeled by the adjoint irrep, see Fig.~\ref{net}.
This simplif\/ication happens because the matrix elements of $L_I^{(\L)}$ can be identif\/ied with the components of the intertwiner~$C^{\L\L'A}$, since these intertwiners  are one-dimensional vector spaces, i.e.
\begin{gather}
\big(L_I^{(\L)}\big)_{\a\b} = C^{\L\L A}_{\a\,\b\,I},\label{geni}
\end{gather}
so that
\begin{gather} \sum_I \big(L_I^{(\L)}\big)_{\a\b}\big(L_I^{(\L')}\big)_{\a'\b'}= \sum_I C^{\L\L A}_{\a\,\b\,\,I}C^{A\L'\L'}_{I\,\a'\,\b'} .\label{heval}\end{gather}
Then the right-hand side of the equation (\ref{heval}) represents the evaluation of the spin network in Fig.~\ref{insertion}.
The condition (\ref{geni}) is not too restrictive since it includes the  $SU(2)$ and $SO(4)$ groups. We need to consider this particular case in order to be able to use the Chain--Mail techniques.

The action of $(\hat V_\s )^n$ in $A_5$ is given by the evaluation of a $\G_{\s,n}$ spin network  which is obtained from the $\G_\s$ spin network by inserting $n$ g-edges labeled by the adjoint irrep. These additional edges connect the edges of $\G_\s$ which correspond to the triangles of the 4-simplex $\s$ where the operators $L_I^{(\L)} \otimes L_I^{(\L')}$ from $\hat V_\s$ act.

Let
\[
Z(M,T)=\sum_{n= 0}^\infty i^n \l^n {Z_n (M,\lambda, T)},
\]
then
\[
Z_0 (M) = Z_0 ({\cal J},M){\Big|}_{{\cal J}=0}.
\]
The state sum $Z_0$ is inf\/inite, unless it is regularized. The usual way of regularization is by using the representations of the quantum group $U_q (\mathfrak g)$ at a root of unity, which, by passing to a f\/inite-dimensional quotient, yields a modular Hopf algebra~\cite{T}. There are only f\/initely many irreps with non-zero quantum dimension in this case, and the corresponding state sum $Z_0$ has the same form as in the Lie group case, except that the usual spin network evaluation used for the spin-foam amplitudes has to be replaced by the quantum spin network evaluation. In this way one obtains a f\/inite and triangulation independent $Z_0$, usually known as the Crane--Yetter invariant~\cite{CY}. This 4-manifold invariant is determined by the signature of the manifold~\cite{Ro,T}. The same procedure of passing to the quantum group at a root of unity  can be applied to the perturbative corrections $Z_n$, but in order to obtain triangulation independent results, the dilute gas limit has to be implemented~\cite{Ba2,MM2}.

\subsection[The Chain-Mail formalism and observables of the Crane-Yetter invariant]{The Chain--Mail formalism and observables of the Crane--Yetter invariant}\label{cmf}

The Chain--Mail formalism for def\/ining the Turaev--Viro invariant and the Crane--Yetter inva\-riant was introduced by Roberts in~\cite{Ro}. In the four-dimensional case, the  construction of the related manifold invariant $Z_0(M)$ had already been implemented by Broda in~\cite{B}. However, the equality  with the Crane--Yetter invariant, as well as the relation of $Z_0(M)$ with the signature of~$M$ appears only in the work of Roberts~\cite{Ro}.

We will follow the conventions of~\cite{BGM}. Let $M$ be a four-dimensional manifold. Suppose we have a handle decomposition \cite{GS,K,RS} of $M$, with a unique 0-handle, and  with $h_i$ handles of order $i$ (where $i=1,2,3,4$).  This gives rise to the link  $\CHL_M$ in the three-sphere $S^3$,  with $h_2+h_1$ components (the ``Chain--Mail link''), which is the Kirby diagram of the handle decomposition~\cite{GS,K}. Namely, we have a dotted unknotted (and 0-framed) circle for each 1-handle of $M$, determining the attaching of the 1-handle along the disjoint union of two balls, and we also have a framed knot for each 2-handle, along which we attach it.
 This is the  four-dimensional  counterpart of the  three-dimensional  Chain--Mail link of Roberts, see~\cite{Ro,BGM}.

The Crane--Yetter invariant $Z_0(M)$, which coincides with the invariant $Z_0(\cj,M)_{\cj=0}$, def\/ined in the introduction, see equation \eqref{CY},   can be represented as a multiple of  the spin-network eva\-lua\-tion of the chain mail link $\CHL_M$, colored with the following linear combination of quantum group irreps (the $\O$-element):
\[
\O = \sum_\L (\dim_q\L)   \L  ,
\]
  see \cite{Ro}. Explicitly, by  using the normalizations of~\cite{BGM}:
\begin{gather}\label{iii}
 Z_0 (M) = \eta^{-\frac{1}{2}(h_2+h_1+h_3-h_4+1) }\langle \CHL_M, \O^{h_2+h_1} \rangle  ,\end{gather}
where
\[
\eta = \sum_\L (\dim_q\L)^2  .
\]
Roberts also proved in \cite{Ro} that  $Z_{0}(M)=\kappa^{s(M)}$, where $\kappa$ is the evaluation of a 1-framed unknot colored with the $\Omega$-element and $s(M)$ denotes the signature of~$M$.

\begin{figure}[t]
\centering
\includegraphics[width=6cm]{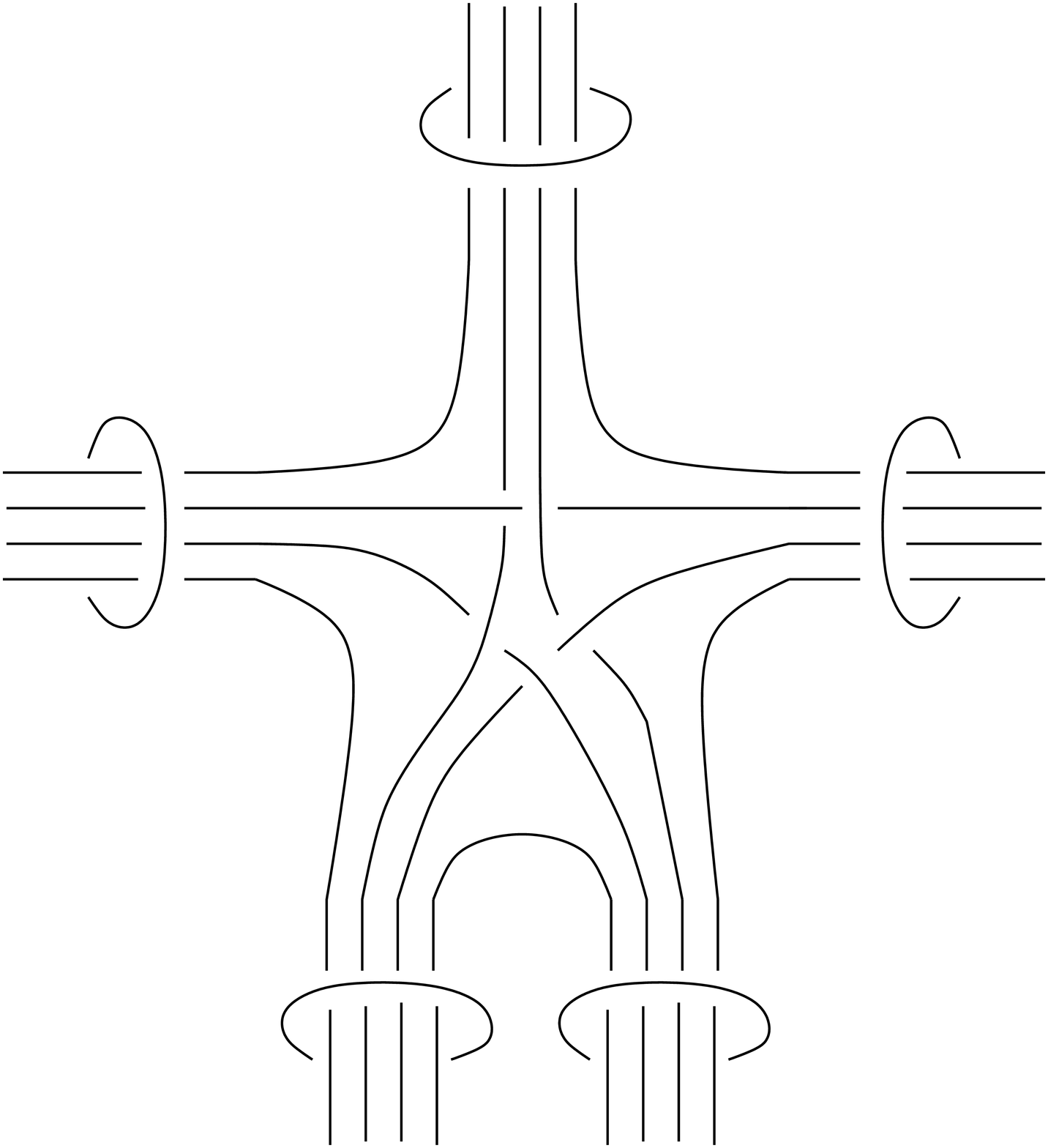}
\caption{Portion of the chain-mail link corresponding to a 4-simplex; this may have additional meridian circles (corresponding to 1-handles) since  we also eliminate pairs of $0$- and $1$-handles, until a single 0-handle is left.}
\label{CHL}
\end{figure}

Given a triangulated manifold $(M,T)$,  consider the natural handle decomposition of $M$ obtained from thickening the dual cell decomposition of $M$; see \cite{Ro,RS}. Then a handle decomposition of $M$ (with a single 0-handle), such that~\eqref{iii} explicitly gives the formula  for $Z_0(M)=Z_0(\cj,M)_{\cj=0}$,  appearing in equation~(\ref{sfgf}),   is obtained from this handle decomposition by canceling pairs of 0- and 1-handles \cite{GS,K,RS}, until a single 0-handle is left; in this case,  in the vicinity of each $4$-simplex, the Chain--Mail link has the form given in Fig.~\ref{CHL}. This   explicit calculation appears in~\cite{Ro,BGM} and  essentially follows from the Lickorish encircling lemma~\cite{KL,L}: the spin-network evaluation of a graph containing a unique strand (colored with  the representation~$\Lambda$) passing through a zero framed unknot colored  with $\Omega$ vanishes unless $\Lambda$ is the trivial representation.

A variant of the Crane--Yetter model $Z_0$ in \eqref{iii} is achieved by inspecting its observables, addressed in~\cite{BGM}. Consider a triangulated 4-manifold~$M$. Consider the handle decomposition of~$M$ obtained from thickening the dual complex of the triangulation, and eliminating pairs of $0$- and $1$-handles until a single 0-handle is left. Any triangle of the triangulation of $M$ therefore yields a 2-handle of~$M$.

{\sloppy Now choose a set  $S$  with $n_S$ triangles of $M$, which will span a (possibly singular) surface $\Sigma_2$ of $M$. Color each  $t \in S$  with  a representation~$\L_t$. The associated observable of the Crane--Yetter invariant is:
\begin{gather}\label{cyo}
Z_0 (M,S,\L_S) = \eta^{-\frac{1}{2}(h_2+h_1+h_3-h_4+1) }\langle \CHL_M; \O^{h_2+h_1-n_S}, \Lambda_S \rangle   \prod_{t \in S} \dim_q(\Lambda_t),
\end{gather}
where $\langle \CHL_M; \O^{h_2+h_1-n_S}, \Lambda_S \rangle $ denotes the spin-network evaluation of the Chain--Mail link~$\CHL_M$, where the components associated with the triangles  $t\in S$ are colored by $\Lambda_s$ and the remaining components with $\O$. We can see ${\rm CHL}_M$ as a pair $(L_S,K_S)$, where $K_S$ denotes the components of the Chain--Mail link given by the triangles of~$S$ and~$L_S$ the remaining components of the Chain--Mail link. We thus have  ${\rm CHL}_M=L_S \cup K_S$.

}

 Let $Z_{\rm WRT}(N,\Gamma)$ denote the Witten--Reshetikhin--Turaev invariant of the colored graph $\Gamma$ embedded in the 3-manifold $N$, in the normalization of \cite{BGM}. Then Theorem~2 of~\cite{BGM} says that:
\begin{gather}\label{cyr}
Z_0 (M,S,\L_S)=Z_{\rm WRT}\big(\partial (M \setminus \hat{\Sigma}_2), K_S'\big)   \kappa^{s (M \setminus \hat{\Sigma}_2)}\,\eta^{\frac{\chi(\Sigma_2)}{2}-{n_S}}  \prod_{t \in S} \dim_q(\Lambda_t).
\end{gather}
Here $\hat{\Sigma}_2$ is an open regular neighborhood of $\Sigma_2$ in $M$, $s$ denotes the signature of the manifold and $\chi$ denotes the Euler characteristic. The link~$ K_S'$ is the link in $\partial (M \setminus \hat{\Sigma}_2)$ along which the 2-handles associated to the triangles of $\Sigma_2$ would be attached, in order to obtain~$M$. This theorem of \cite{BGM} essentially follows from the fact that the pair $(L_S,K_S)$ is a surgery presentation of the pair $\big(\partial (M \setminus \hat{\Sigma}_2), K_S'\big)$, a link embedded in a manifold,  apart from connected sums with~$S^1 \times S^2$.

\section{The f\/irst-order correction}\label{foc}

Recall that there are two possible ways of representing the Crane--Yetter invariant $Z_0$: as a state sum invariant \eqref{CY} and as the evaluation of a Chain--Mail link~\eqref{iii}. It follows from~(\ref{osfpi}) that~$Z_1$  can be written as $NZ'_0$ where~$N$ is the number of 4-simplices of $T(M)$ and $Z'_0$ is the state sum given by a modif\/ication of the state sum $Z_0$ where a single 4-simplex is perturbed by the operator~$\hat V_\s$.

In order to calculate $Z_0'$ consider a 4-manifold $M$  with a triangulation $T$ whose dual complex is $T^*$. Given a 4-simplex $\s\in T$ we def\/ine an insertion $\I $ as being a choice of a pair of triangles of $\s$ which do not belong to the same tetrahedron of~$\s$ and have therefore a single vertex in common (following~\eqref{sfpi} we will distinguish the order in which the triangles appear). Given  the colorings~$\L_f$  of the triangles of~$\s$ {(or of the dual faces $f$)} and  the colorings $\iota_l$ of the tetrahedrons of $\s$ {(or of the dual edges $l$)}, then $A_5(\L_f ,\iota_l ,\I)$ is the evaluation of the spin network of Fig.~\ref{net}. We then have:
\begin{gather}\label{z0p}
Z_1(M,T) = \frac{1}{5}\sum_{ \s}   \sum_{\I_\s}  \sum_{\L_f, \iota_l} A_5 (\L_{f(\s)} , \iota_{l(\s)} , \I_\s) \prod_f \dim \L_f  \prod_{v\ne v(\s)}  A_5(\L_{f(v)}, \iota_{l(v)}) ,\end{gather}
where $v(\s)$ is the vertex of $T^*$ corresponding to $\s$. This sum is over the set of all 4-simplices $\s$ of~$T(M)$, as well as over the set $\I_\s$ of all insertions  of~$\s$ and over the set of all colorings $(\L_f, \iota_l)$ of the faces and the edges of $T^*(M)$ (or equivalently, a sum over the colorings of the triangles and the tetrahedrons of $T(M)$.)

The inf\/inite sum in~(\ref{z0p})  is regularized by passing to the category of representations of the quantum group $U_q (g)$, where $q$ is a root of unity. In order to calculate $Z_1(M,T)$ in this case, let us represent it as an evaluation of the Chain--Mail link $\CHL(M,T)$~\cite{Ro} in the three-sphere~$S^3$.

\begin{figure}[t]
\centering
\includegraphics{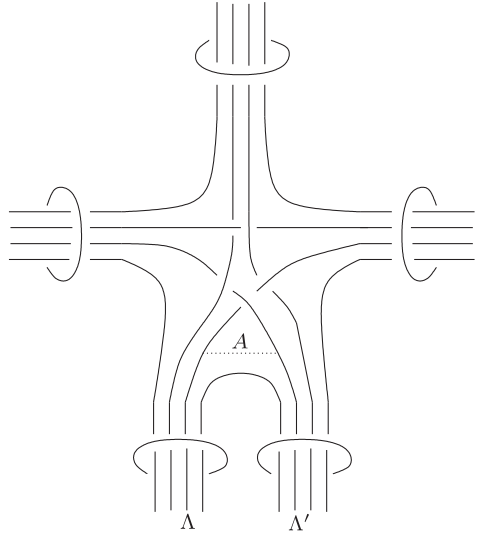}
\caption{Portion of the graph $CML_M^\I$ corresponding to a 4-simplex with an insertion. All strands are to be colored with $\Omega$, unless they intersect the insertion, as indicated.}
\label{CI}
\end{figure}

As  explained in Subsection~\ref{cmf}, the invariant $Z_0(M)$ can be represented as a multiple of  the evaluation of the chain mail link $\CHL_M$ colored with the linear combination of {the quantum group} irreps
$\O = \sum_\L (\dim_q\L)   \L $, see equation \eqref{iii}. Analogously,  by extending the 3-dimensional approach of~\cite{MM2}, a chain-mail formulation for the equation (\ref{z0p}) can be given.   Consider the  handle decomposition of~$M$ obtained by thickening the dual cell decomposition of~$M$ associated to the triangulation $T$ of $M$. One can cancel pairs of 0- and 1-handles, until a~single 0-handle is left. Let $\CHL_M$ be the associated chain-mail link (the Kirby diagram of the handle decomposition). We then have
\begin{gather}
 Z_1 (M,T)= {\frac{1}{5}} \sum_{ \I}  \sum_{\L,\L'} \eta^{-\frac{1}{2}(h_2+h_1+h_3-h_4+1) }   \dim_q\L   \dim_q\L'  \langle \CHL_M^\I, \O^{h_2+h_1-2}, \L, \L' \rangle,  \label{obs}
\end{gather}
where, as before, an insertion $\I$ is the choice of a pair of triangles $t_1$ and $t_2$ in some 4-simplex of $M$, such that $t_1$ and $t_2$ have only one vertex in common. Given an insertion $\I$, the graph $\CHL_M^\I$ is obtained from the chain-mail link $\CHL_M$ by inserting a single edge (colored with the adjoint representation of $\mathfrak g$) connecting the components of $\CHL_M$ (colored with $\L$ and $\L'$) corresponding to $t_1$ and $t_2$; see Fig.~\ref{CI}. $\CHL_M^\I$ can be considered as a pair $(L_\I,\G_\I)$
where $L_\I$ denotes the components of $\CHL_M^\I$ not incident to the insertion  $\I$ (which are exactly $h_2+h_2-2$) and  $\G_\I$ denotes the component of $\CHL_M^\I$ containing the insertion $\I$. Hence we use the notation  $\langle \CHL_M^\I, \O^{h_2+h_1-2}, \L, \L' \rangle$ to mean the evaluation of the pair $(L_\I,\G_\I)$ where all components of~$L_\I$ are colored with $\Omega$ and the two circles of $\G_\I$ are colored with $\L$ and $\L'$, with an extra edge connecting them, colored with the adjoint representation $A$.

\begin{figure}[t]
\centering
\includegraphics{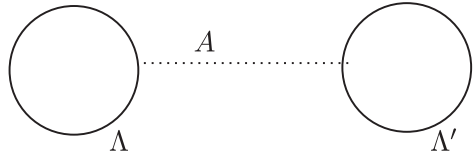}
\caption{Spin network $\Gamma_1(\L,\L')$. Here $A$ is the adjoint representation.}
\label{Lambda}
\end{figure}

Consider an insertion $\I$ connecting the triangles $t_1$ and $t_2$, which intersect at a single vertex. Equation~(\ref{obs}) coincides, apart from the inclusion of the single insertion, with the observables of the Crane--Yetter invariant \cite{BGM} (Subsection~\ref{cmf}) for the pair of triangles colored with~$\L$ and~$\L'$; see equation~\eqref{cyo}.  By using the discussion in Section~\ref{section5} and Theorem 2 of~\cite{BGM} (see equation~\eqref{cyr}) one therefore obtains, for each insertion $\I$ and each pair of irreps $\L$, $\L'$:
\begin{gather}\label{CC} \eta^{-\frac{1}{2}(h_2+h_1+h_3-h_4+1) }\langle \CHL_M^\I, \O^{h_2+h_1-2}, \L, \L' \rangle = Z_{\rm WRT}(S^3,\Gamma(\L,\L')) \eta^{-\frac{3}{2}}Z_0(M),
\end{gather}
where $\Gamma(\L,\L')$ is the colored link of Fig.~\ref{Lambda}.  In addition $Z_{\rm WRT}$ denotes the Witten--Reshetikhin--Turaev  invariant~\cite{W,RT} of graphs in manifolds, in the normalization of \cite{BGM,MM}. Note that in the notation of equation~\eqref{cyr}, $\Sigma_2=t_1 \cup t_2$ is two triangles which intersect at a vertex, thus $\chi(\Sigma_2)=1$ and also its regular neighborhood $\hat{\Sigma}_2$ is homeomorphic to the 4-disk, thus $s(M)=s(M \setminus \hat{\Sigma}_2)$.

Equation (\ref{CC}) follows essentially from the fact that
the pair $\CHL_M^\I=(L_\I,\G_\I)$
 is a surgery presentation~\cite{K,GS} of the pair $(S^3,\Gamma(\L,\L'))$, apart from connected sums with $S^1 \times S^2$;  c.f.\ Theorem~\ref{obsA}, below.

To see this, note that  (after turning the circles associated with the 1-handles of $M$ into dotted circles) the link $L_\I$ is a Kirby diagram for the manifold $M$ minus an open regular neighborhood~$\Sigma'$ of the 2-complex $\Sigma$ made from the vertices and edges of the triangulation of~$M$, together with the triangles $t_1$ and $t_2$.  Since $t_1$ and $t_2$ intersect at a single vertex,  any regular neighborhood $\hat{\Sigma}_2$ of the (singular) surface $\Sigma_2$ spanned by $t_1$ and $t_2$ is homeomorphic to the 4-disk $D^4$. Therefore  $\Sigma'$ is certainly homeomorphic to the boundary connected sum  $\big( \natural_{i=1}^k (D^3 \times S^1)\big) \natural D^4,$ whose boundary is $\big (\#_{i=1}^k (S^2 \times S^1) \big ) \# S^3$, for some positive integer $k$. Here $\#$ denotes the connected sum of manifolds and $\natural$ denotes the boundary connected sum of manifolds. The circles {$c_1,c_2\subset \Gamma_\I$} associated with the triangles $t_1$ and $t_2$ def\/ine a link which lives in $\partial \hat{\Sigma}_2\cong S^3 \subset \big(\#_{i=1}^k (S^2 \times S^1) \big) \# S^3$.

\begin{figure}[t]
\centering
\includegraphics{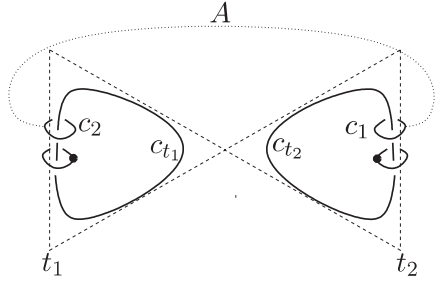}
\caption{The Kirby diagram for $P$ in the vicinity of the triangles $t_1$ and $t_2$. We show the belt-spheres~$c_1$ and~$c_2$ of the 2-handles of $P$ (attaching along~$c_{t_1}$ and~$c_{t_2}$) associated with the triangles~$t_1$ and~$t_2$.}
\label{arg}
\end{figure}

The two circles  $c_1$ and $c_2$  def\/ine a 0-framed unlink in $S^3$, with each individual component being unknotted.
Let us see why this is the case.  We will turn the underlying handle decomposition of $M$ upside down, by passing to the dual handle decomposition of $M$, where each $i$-simplex of the triangulation of $M$ yields an $i$-handle of $M$; see~\cite[p.~107]{GS}. Consider the bit $P\subset M$ of the handle-body  yielded by the 2-complex $\Sigma$, thus $P$ is (like $\Sigma'$) a regular neighborhood of~$\Sigma$.  Maintaining the 0-handle generated by  the vertex $t_1 \cap t_2$, eliminate some pairs of $0$- and $1$-handles, in the usual way, until a single $0$-handle of $P$ is left. Clearly $\partial P^*=\partial (M \setminus \Sigma')$, where~$*$  denotes the orientation reversal. The circles $c_1$ and $c_2$, in $\partial P^*$, correspond now (since we considered the  dual handle decomposition) to the belt-spheres of the 2-handles of $P$ (attached along~$c_{t_1}$ and~$c_{t_2}$) and associated with the triangles~$t_1$ and~$t_2$. Since $c_1$ and $c_2$ are 0-framed meridians going around~$c_{t_1}$ and~$c_{t_2}$  (see~\cite[Example 1.6.3]{GS}) it therefore follows that these circles are unlinked and are also, individually,  unknotted; see Fig.~\ref{arg}. Given this and the fact that the insertion $\I$ colored with $A$ also lives in $S^3$, it follows that   $\CHL_M^\I=(L_\I,\G_\I)$
is a surgery presentation of the pair $(S^3,\Gamma(\L,\L'))$, apart from the connected sums, distant from $\G(\L,\L')$, with $S^1 \times S^2$.

Since the evaluation of the tadpole spin network is zero, it follows that  $Z_{\rm WRT}(S^3,\Gamma(\L,\L')){=0}$, and consequently

\begin{Theorem}\label{firstorder}For any triangulation $T$ of $M$ we have $ Z_1(M,T) =0.$
 \end{Theorem}

\section{The second-order correction}\label{soc}

Since $Z_1=0$, we have to calculate $Z_2$ in an appropriate limit such that the partition function~$Z$ is dif\/ferent from $Z_0$ and such that $Z$ is independent of the triangulation \cite{MM2}. Let $N$ be the number of $4$-simplices.  From (\ref{osfpi}) we obtain
\begin{gather}
Z_2(M) = \frac{1}{2}\sum_{\L_f, \iota_l}\prod_f \dim \L_f \left( \sum_\s \hat V^2_\s +\sum_{\s\ne\s'}\hat V_\s \hat V_{\s'}\right)\prod_v A_5 (\L_{f(v)} , \iota_{l(v)}),\label{ztwo}
\end{gather}
where
\begin{gather}\label{vtoc}
{\hat V_{\s} A_5 (\L_{f(v)} , \iota_{l(v)})=\frac{1}{5}\sum_{\textrm{insertions } \I \textrm{ of } \s}  A_5 (\L_{f(v)} , \iota_{l(v)},\I)} ,
\end{gather}
if $\s$ is dual to $v$, see Fig.~\ref{net}. On the other hand, $\hat{V_\s} A_5 (\L_{f(v)} , \iota_{l(v)})=A_5 (\L_{f(v)} , \iota_{l(v)})$ if $v$ is not dual to $\s$. In order to to solve the possible framing and crossing ambiguities arising from the equation (\ref{ztwo}), a method analogous to the one used in \cite{MM2} can be employed. Note that there are exactly 30 insertions in a $4$-simplex $\s$, corresponding to pairs of triangles of $\s$ with a single vertex in common. This is because there are exactly three triangles of $\sigma$ having only one vertex in common with a given triangle of $\s$.

Analogously to the f\/irst-order correction, ${Z_2}$ can be written as
\begin{gather} {Z_2} (M,T) =\frac{1}{2}\eta^{-\frac{1}{2}(h_2+h_1+h_3-h_4+1)} \sum_{k=1}^N \langle \CHL_M, \O^{h_2+h_1}, \hat V_k^2 \rangle\nonumber\\ \phantom{{Z_2} (M,T) =}{}
+\frac{1}{2}\eta^{-\frac{1}{2}(h_2+h_1+h_3-h_4+1)} \sum_{1\le k\ne l \le N}\langle \CHL_M, \O^{h_2+h_2}, \hat V_k ,\hat V_l \rangle , \label{AA}
\end{gather}
where the f\/irst sum denotes the contributions from two insertions $\hat{V}$ in the same 4-simplex~$\s_k$ and the second sum represents the contributions when the two insertions $\hat{V}$ act in dif\/ferent 4-simplices~$\s_k$ and~$\s_l$. As in the previous section, we will use the handle decomposition of~$M$ with an unique 0-handle naturally obtained from the thickening of~$T^*(M)$.

Note that each $\langle \CHL_M, \O^{h_2+h_1}, \hat V_k^2 \rangle$ corresponds to a sum over all the possible choices of pairs of insertions in the 4-simplex $\s_k$. The value of $\langle \CHL_M, \O^{h_2+h_1}, \hat V_k^2 \rangle$ is obtained from the evaluation of the chain-mail link $\CHL_M$ colored with~$\Omega$,  which contains g-edges carrying the adjoint representation, as in the calculation of the f\/irst-order correction.

A conf\/iguration $C $ is, by def\/inition, a choice of insertions distributed along a set  of 4-simplices of $M$. Given a positive integer $n$ and a set $R$ of 4-simplices of $M$,  we denote by $ {\cal C}^n_R$ the set of conf\/igurations with $n$ insertions distributed along~$R$. By expanding each $V$ into a sum of insertions, the equation (\ref{AA}) can be written as:
\begin{gather}
{Z_2} (M,T) =\frac{1}{50}\eta^{-\frac{1}{2}(h_2+h_1+h_3-h_4+1)} \sum_{k=1}^N \sum_{C \in {\cal C}^2_{{\{\s_k\}}}} \langle \CHL_M^C, \O^{h_2+h_1} \rangle
\nonumber\\
\phantom{{Z_2} (M,T) =}{}+\frac{1}{50}\eta^{-\frac{1}{2}(h_2+h_1+h_3-h_4+1)}  \sum_{1\le k\ne l \le N} \sum_{C \in {\cal C}^2_{{\{\s_k, \s_l\}}}}  \langle \CHL_M^C, \O^{h_2+h_2} \rangle . \label{AAA}
\end{gather}
Note that each graph $\CHL_M^C$ splits naturally as $(L_C,\G_C)$, where the f\/irst component contains the circles non incident to any insertion of $C$.

\begin{figure}[t]
\centering
\includegraphics{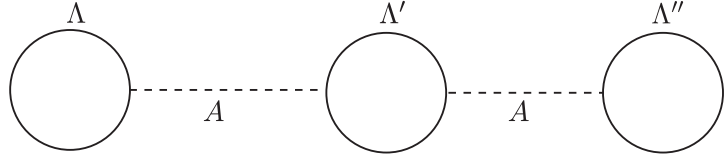}
\caption{Colored graph $\Gamma_2'(\L,\L',\L'')$, a wedge graph. Here $A$ is the adjoint representation.}
\label{wedge}
\end{figure}

The second sum in equation (\ref{AAA}) vanishes, because it is the sum of terms proportional to:
\begin{gather}\label{BB}
\left (\sum_{\L,\L'} \langle \G_1(\L,\L')\rangle\right)^2  Z_0 (M)
\end{gather}
and to
\begin{gather}\label{3b}\sum_{\L,\L',\L''} \langle \G_2'(\L,\L',\L'')\rangle Z_0(M) ,
 \end{gather}
where $\G_1$ is the dumbbell  spin network of Fig.~\ref{Lambda}, and $\G_2'$ is a three-loop spin network, see Fig.~\ref{wedge}.

 These spin networks arise from  the cases when the  pair   $(L_C,\G_C)$
is a surgery presentation of {the disjoint union} $\left(S^3,\G_1\right)\sqcup \left(S^3,\G_1\right)$ and of $\left(S^3, \G_2'\right)$, respectively, apart from connected sums with $S^1\times S^2$; see Theorem~\ref{obsA} below.
The former case corresponds to a situation where the two insertions act in pairs of triangles without a common triangle, and the latter corresponds to a situation where the  two pairs of triangles have a triangle in common, which necessarily is a~triangle in the intersection $\s_k \cap \s_l$ of the 4-simplices~$\s_k$ and~$\s_l$. The evaluations in~(\ref{BB}) and~(\ref{3b}) vanish since the corresponding spin networks have  tadpole subdiagrams.

The f\/irst sum in~(\ref{AAA})  also gives  the terms  proportional to the ones in equations~(\ref{BB}) and~(\ref{3b}). These terms correspond  to two  insertions connecting two pairs made from four distinct triangles of~$\s_k$ and to two insertions connecting two pairs of triangles made from  three distinct triangles of~$\s_k$, respectively. All these terms vanish.

\begin{figure}[t]
\centering
\includegraphics{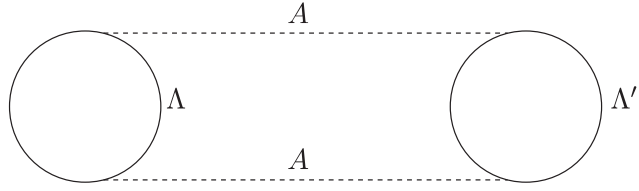}

\caption{Colored graph $\Gamma_2(\L,\L')$, a double dumbbell graph. As usual $A$ is the adjoint representation.}
\label{dumbbell2}
\end{figure}

The non-vanishing terms in equation (\ref{AAA}) arise from a pair of insertions connecting the same two triangles in a 4-simplex. There are exactly 30 of these. Therefore, by using Theorem~\ref{obsA} of Section~\ref{five}, we obtain:
\begin{gather*}
Z_2(M,T) = {{\frac{3N}{5}\eta^{-2}}} \sum_{\L,\L'}\dim_q\L  \dim_q\L'   \langle \G_2 (\L,\L')\rangle Z_0 (M),
\end{gather*}
where $\G_2$ is a two-handle dumbbell spin network, see Fig.~\ref{dumbbell2}.
We thus have:
\begin{Theorem}\label{qwer}
The second-order perturbative correction $Z_2(M,T)$ divided by the number of $N$ of $4$-simplices of the manifold is  triangulation independent. In fact:
\[
\frac{Z_2(M,T)}{N} = {{\frac{3}{5}\eta^{-2}}} \sum_{\L,\L'}\dim_q(\L) \dim_q(\L')   \langle \G_2 (\L,\L') \rangle Z_0 (M).
\]
\end{Theorem}

Here $\langle \G_2 (\L,\L') \rangle$ denotes the spin-network evaluation of the colored graph $ \G_2 (\L,\L')$. Note that
\begin{gather*}
\langle \G_2 (\L,\L')\rangle = \frac{\theta(A,\L,\L) \theta(A,\L',\L')}{ \dim_q A}   ,
\end{gather*}
which is is obviously non-zero, and therefore $Z_2(M,T) \neq 0$.

\section{Higher-order corrections}\label{five}

For $n>2$, the contributions to the partition function will be of the form
 \begin{gather*}
\eta^{-\frac{1}{2}(h_2+h_1+h_3-h_4+1)}\langle \CHL_M , \O^{h_1+h_2}, (\hat V_1)^{k_1}\cdots (\hat V_N)^{k_N}\rangle ,
\end{gather*}
where $k_1+\cdots +k_N=n$.
By using the equation (\ref{vtoc}), each of these terms splits as a sum of terms of the form:
\begin{gather}\label{eee}
\frac{1}{5^n}\eta^{-\frac{1}{2}(h_2+h_1+h_3-h_4+1)}\langle \CHL_M^{C} , \O^{h_1+h_2}\rangle=\langle C\rangle,
\end{gather}
where $C$ is a set of $n$ insertions (a ``conf\/iguration'') distributed among the $N$ 4-simplices of the chosen triangulation of $M$, such that the 4-simplex $\s_i$ has $k_i$ insertions. Insertions are added to the chain-mail link $\CHL_M$ as in Fig.~\ref{CI}, forming a graph $\CHL_M^C$ (for framing and crossing ambiguities we refer to \cite{MM2}).

Note that each graph $\CHL_M^C$ splits as $(L_C,\G_C)$ where $L_C$ contains the components of  $\CHL_M$ not incident to any insertion. As in the $n=1$ and $n=2$ cases, equation (\ref{eee}) coincides, apart from the extra insertions, with the observables of the Crane--Yetter invariant def\/ined in~\cite{BGM}; see Subsection~\ref{cmf}. Therefore, by using the same argument that proves Theorem 2 of~\cite{BGM} we have:

\begin{Theorem}\label{obsA} Given a configuration  $C$ consider the $2$-complex $\Sigma_C$ spanned by the $k$ triangles of~$M$ incident to the insertions of~$C$. Let $\Sigma'_C$ be  a regular neighborhood of $\Sigma_C$ in $M$. Then~$M$ can be obtained from  $M \setminus \Sigma'_C$ by adding the $2$-handles corresponding to the faces of $\Sigma$ $($and some further $3$- and $4$-handles, corresponding to the edges and vertices of $\Sigma)$. These $2$-handles attach along a framed link $K$  in $\partial( M \setminus \Sigma')$, a manifold diffeomorphic to  $\partial(\Sigma_C')$ with the reverse orientation. The insertions of $C$ can be transported to this link $K$ defining a graph $K_C$ in $\partial(\Sigma'^*_C)=\partial(M \setminus \Sigma_C')$. We have:
\begin{gather*}
\langle C\rangle =\sum_{\L_1,\ldots , \L_k} Z_{\rm WRT}\big(\partial(\Sigma_C'^*), K_C; \L_1,\ldots,\L_k\big)\kappa^{s(M \setminus \Sigma')}\eta^{\frac{\chi(\Sigma)}{2} - k}\dim_q\L_1 \cdots \dim_q \L_k,
\end{gather*}
where $s(M \setminus \Sigma')$ denotes the signature of the manifold $M \setminus \Sigma'$ and $\chi$ denotes the Euler characteristic.
\end{Theorem}

Note that, up to connected sums with $S^1 \times S^2$, the pair $\CHL_M^C=(L_C,\G_C)$ is a surgery presentation of $(\partial(M \setminus \Sigma_C'), K_C)=(\partial(\Sigma_C'^*),K_C).$
Unlike the $n=1$ and $n=2$ cases, it is not possible to determine the pair $(\partial(\Sigma_C'^*),K_C)$ for $n\ge 3$ without having an additional information about the conf\/iguration. In fact, considering the set of all triangulations of $M$, an inf\/inite number of dif\/feomorphism classes for $(M \setminus \Sigma', K_C)$ is in general possible for a f\/ixed $n$; see~\cite{MM2} for the three dimensional case. This makes it complicated to analyze the triangulation independence of the formula for~$Z_n(M,T)$ for  $n\ge 3.$

Since
\[
Z(M,T) = \sum_n {\l^n } Z_n(M,T),
\]
where
\begin{gather*}
Z_n(M,T)=  \sum_{{k_1+\cdots+k_N=n}} \frac{1}{ k_1!\cdots k_N!}\eta^{-\frac{1}{2}(h_2+h_1+h_3-h_4+1)}  \langle \CHL_M , \O^{h_1+h_2}, (\hat V_1)^{k_1}\cdots (\hat V_N)^{k_N}\rangle,
\end{gather*}
in order to resolve the triangulation dependence of $Z_n$, let us introduce the quantities
\begin{gather}\label{dilute}
z_n = \lim_{N\to\infty} \frac{Z_n (M,T)}{ N^{\frac{n}{2}} Z_0(M)} ;
\end{gather}
see \cite{MM2,Ba2}. The limit is to be extended to the set of all triangulations $T$ of $M$, with  $N$ being the number of 4-simplices of $M$, in a sense  to be made precise; see \cite{MM2}. From Sections~\ref{foc},~\ref{soc} and Theorem~\ref{qwer} it follows that
\begin{gather*}
 z_1=0   , \qquad z_2 = {\frac{3\eta^{-2}}{5\dim_q A}}\left(\sum_{\L}\dim_q \L \theta(A,\L,\L) \right)^2  .
 \end{gather*}
Note that the values of $z_1$ and $z_2$ are universal for all compact 4-manifolds.
The expression for~$z_2$ is f\/inite because there are only f\/initely many irreps for the quantum group $U_q (\mathfrak g)$, of non-zero quantum dimension, when $q$ is a root of unity.

\section{Dilute-gas limit}\label{section5}

We will now show how to def\/ine and calculate the limit in the equation~(\ref{dilute}). Let $M$ be a~4-manifold and let us consider a set $\cal{S}$ of triangulations of $M$, such that for any given $\epsilon >0$ there exists a triangulation $T \in \cal{S}$  such that the diameter of the biggest 4-simplex is smaller than $\epsilon$, i.e.\ the triangulations in ${\cal S}$ can be chosen to be arbitrarily f\/ine. We want to calculate the limit in equation~(\ref{dilute}) only  for triangulations belonging to the set ${\cal S}$.

Furthermore, we suppose that $\cal{S}$ is such that (c.f.~\cite{MM2}):
\begin{Restriction}[Control of local complexity-I]\label{A1} Together with the fact that the triangulations in~${\cal S }$ are arbitrarily fine we suppose that:

 There exists a positive integer $L$ such that any $4$-simplex of any triangulation $T \in \cal{S}$ intersects at most $L$ $4$-simplices of $T$.
\end{Restriction}

Let us f\/ix $n$ and  consider $Z_{2n}(M,T)$ when $N\to\infty$.  The value of $Z_{2n}$ will be given as a sum of contributions of conf\/igurations ${C}$ such that $n_1$ insertions  of $\hat V$ act in a 4-simplex $\s_1$, $n_2$ of insertions of $\hat V$ act in the 4-simplex $\s_2 \ne \s_1$ and so on,  such that $n_1 + n_2 + \cdots+ n_N=2n$ and $n_k \geq 0$.

A conf\/iguration for which any 4-simplex has either zero or two insertions, with all 4-simplices which have insertions being disjoint will be called a dilute-gas conf\/iguration. There will be
\[
15^n{N!\over n!(N-n)!}- \delta (N,T) ,
\]
dilute-gas conf\/igurations, where $\delta$ is the number of pairs of 4-simplices in $T$ with non-empty intersection. From  Restriction~\ref{A1} it follows that $\delta(N,T) = O(N)$ as $N\to\infty$.

Each dilute-gas conf\/iguration contributes $z_2^n  Z_0 (M)$ to $Z_{2n}(M,T)$
and we can write
\begin{gather} \label{DDD}\frac{Z_{2n}}{ Z_0} = \left({N!\over n!(N-n)!}- O(N)\right) z_2^n  + \sum_{\textrm{non-dilute } C} {\langle C \rangle\over Z_0} ,\end{gather}
where $\langle C\rangle=\frac{1}{5^n}\eta^{-\frac{1}{2}(h_2+h_1+h_3-h_4+1)}\langle \CHL_M^{C} , \O^{h_1+h_2}\rangle,$ denotes the contribution of the conf\/igu\-ra\-tion~$C$.

Let us describe the contribution of the  non-dilute conf\/igurations $C$ more precisely.  Recall that a conf\/iguration $C$ is given by  a choice of $n$ insertions connecting $n$ pairs of triangles of $M$, where each pair belongs to the same 4-simplex of $M$ and the triangles have only one common vertex.

Given a conf\/iguration $C$ with $n$ insertions, consider  a (combinatorial) graph $\gamma_C$ with a vertex for each triangle appearing in $C$ and for each insertion and edge connecting the corresponding vertices. The graph $\g_C$ is obtained from $\G_C$ by collapsing the circles of $\G_C$ of it into vertices.  However note that $\g_C$ is merely a combinatorial graph, whereas $\G_C$ is a graph in $S^3$, which can have a complicated embedding.

If $\g_C$ has a connected component homeomorphic to the graph made from two vertices and an edge connecting them, then  $\langle C\rangle$ vanishes, since in this case the embedded graph whose surgery presentation is given by $(\CHL_M^C,\G_C)$ will have a tadpole. In fact, looking at Theorem~\ref{obsA}, one of the connected components of $(\partial(\Sigma'^*),K_C)$ will be $(S^3,\G_1) $, where $\G_1$ is the graph in Fig.~\ref{Lambda}.

Consider a manifold with a triangulation with $N$ 4-simplices, satisfying Restriction~\ref{A1}. The number of possible conf\/igurations $C$ with $l$ insertions with make a connected graph $\g_{C}$ is bounded by $N(10L)^{l-1}(l-1)!$. In particular the number of non-dilute conf\/igurations $V$ with $2n$ insertions and yielding a non-zero contribution is bounded by
\begin{gather}\label{lll} \max_{\substack{l_1+\cdots + l_{2n}=2n\\ l_i\neq 1\\ \exists_i \colon l_i \ge 3}} b \prod_{i=1}^{2n}N(10L)^{l_k-1}(l_k-1)!=O\big(N^{n-1}\big).  \end{gather}
This is simply the statement that if a graph $\g_C$ has $k$ connected components, then it has $O(N^k)$ possible conf\/igurations. Since $k\le n-1$ for a non-dilute conf\/iguration, the bound~(\ref{lll}) follows.

We now need to estimate the value of $\langle C \rangle$ for a non-dilute conf\/iguration $C$. We will need to make the following restriction on the set  ${\cal S}$. We refer to the notation introduced in Theorem \ref{obsA}.

\begin{Restriction}[Control of local complexity-II]\label{A2}
The set of ${\cal S}$ of  triangulations of $M$ is such that given a positive integer $n$ {then the} number of possible diffeomorphism classes for the pair $(\partial(\Sigma_C'),K_C)$ is finite as we vary the triangulation $T \in {\cal S}$ and the configuration $C$ with $n$ insertions.
\end{Restriction}

In the three-dimensional case  a class of triangulations ${\cal S}$ satisfying Restrictions \ref{A1} and \ref{A2} was constructed by using a particular class of cubulations  of  3-manifolds (which always exist, see~\cite{CT}) and their barycentric subdivisions, see \cite{MM2}.  These cubulations have a simple local structure, with only three possible types of local conf\/igurations, which permits a case-by-case analysis as the cubulations are ref\/ined through barycentric subdivisions. In the case of four-dimensional cubulations, no such list is known, although it has been proven that a f\/inite (and probably huge) list exists. Therefore the approach used in the three-dimensional case cannot be directly applied to the four-dimensional case. However, it is reasonable to assume that triangulations coming from the  barycentric subdivisons of a cubulation of $M$ satisfy Restriction~\ref{A2}.

More precisely, given a cubulation $\square$ of $M$, let $\Delta_\square$ be the triangulation obtained from $\square$ by taking the cone of each $i$-face of each cube of $\square$, starting with the 2-dimensional faces. Consider the class ${\cal S}=\{\Delta_{\square^{(n)}}\}_{n=0}^\infty$, where $\square^{(n)}$ is the barycentric subdivision of order~$n$ of $\square$. Then we can see that Restriction~\ref{A1} is satisf\/ied by this example, and we conjecture that ${\cal S}$ {also} satisf\/ies Restriction~\ref{A2}.

The Restriction~\ref{A2} combined with Theorem~\ref{obsA} implies that the value of $\langle C\rangle$ in equation~\eqref{DDD} is bounded for a f\/ixed $n$, considering the set of all triangulations in the set ${\cal S}$ and all possible conf\/igurations with $2n$ insertions. {Since  the number of non-dilute conf\/igurations $C$ which have a non-zero contribution is of {$O(N^ {n-1})$,} it}  follows that
\[
\sum_{\textrm{non-dilute } C} {\langle C \rangle\over Z_0} = O\big(N^{n-1}\big)  .
\]
Therefore
\begin{gather*}
\lim_{N\to\infty} \frac{Z_{2n} (M,T) }{N^{{n}} Z_0(M)}= \frac{z_2^n}{n!} ,
\end{gather*}
or
\begin{gather}\label{zeva}
\frac{Z_{2n} (M,T)}{ Z_0(M)}\approx \frac{z_2^n}{n!} N^n  ,
\end{gather}
for large $N$.

\begin{figure}[t]
\centering
\includegraphics{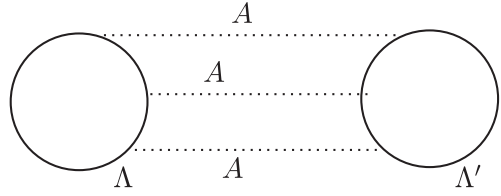}
\caption{Colored graph $\Gamma_3(\L,\L')$, a triple dumbbell graph. As usual $A$ is the adjoint representation.}
\label{dumbbel3}
\end{figure}

\begin{figure}[t]
\centering
\includegraphics{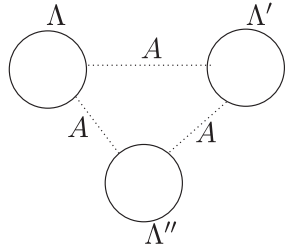}
\caption{Colored graph $\Gamma_3'(\L,\L',\L'')$. As usual $A$ is the adjoint representation.}
\label{refer}
\end{figure}

In the case of $Z_{2n+3}$, the dominant conf\/igurations for triangulations with a large number $N$ of 4-simplices consist of conf\/igurations $C$ (as before called dilute)  whose associated combinatorial graph $\g_C$ has as connected components a  connected closed graph with three edges  and $(n-1)$ connected closed graphs with two  edges.  We can write:
\begin{gather*}{Z_{2n+3}\over Z_0} = \sum_{\textrm{dilute } C} \langle C \rangle +\sum_{\textrm{non-dilute } C} \langle C \rangle  .
\end{gather*}
Since the number of dilute conf\/igurations is {of} {$O(N^ n)$,} while the second sum is {of} $O(N^{n-1})$, due to the Restrictions~\ref{A1} and~\ref{A2}, we then obtain for large $N$
\begin{gather*}{Z_{2n+3}\over Z_0} = O(N^{n}) .
\end{gather*}
More precisely
\[
{Z_{2n+3}\over Z_0} \approx z_3 (z_2)^n N\frac{(N-1)!}{n!(N-1-n)!} ,
\]
or
\begin{gather}
{Z_{2n+3}\over Z_0} \approx z_3 (z_2)^n \frac{N^n}{n!} ,\label{zoda}
\end{gather}
for large $N$, where $z_3$ is the sum of two terms. The f\/irst term is
\[
\frac{30}{6 \times 5^3}
\sum_{\Lambda,\Lambda'} \dim_q \L   \dim_q \L'    \langle \Gamma_3 (\L,\L')\rangle ,
 \]
 where $\Gamma_3$ is the triple dumbbell graph of Fig.~\ref{dumbbel3}, corresponding to three insertions connecting the same pair of triangles of the underlying 4-simplex (there are exactly 30 of these). The second term is
\[
\frac{30}{6 \times 5^3}
\sum_{\Lambda,\Lambda',\Lambda''} \dim_q \L  \dim_q \L'   \dim_q \L'' \langle \Gamma_3' (\L,\L',\L'')\rangle ,
\]
where $\G_3'$ appears in Fig.~\ref{refer}. This corresponds to three insertions making a chain of triangles, pairwise having only  a vertex in common (there are exactly 30 insertions like these for each $4$-simplex).

\subsection[Large-$N$ asymptotics]{Large-$\boldsymbol{N}$ asymptotics}

Let us now study the asymptotics of $Z(M,T)$ for $N\to\infty$. We will denote $Z(M,T)$ as $ Z(\l,N)$ and $Z_0(M)$ as $Z_0 (\l_0)$, in order to highlight the fact that the Crane--Yetter state sum $Z_0 (M)$ can be understood as a path integral for the $BF$-theory with a cosmological constant term $\l_0 \int_M \langle B\wedge B\rangle$, such that $\l_0 $ is a certain function of an integer~$k_0$, which specif\/ies the quantum group at a root of unity whose representations are used to construct the Crane--Yetter state sum. In the case of a quantum $SU(2)$ group it has been conjectured that $\l_0 = 4\pi /k_0$, see~\cite{smolin}. Consequently
\begin{gather*}
Z(\l)  = \int {\cal D}A\,{\cal  D}B\, e^{i\int_M \langle B\wedge F + \lambda  B\wedge B \rangle}
 =  \int {\cal D}A {\cal  D}B  e^{i\int_M \langle B\wedge F + \lambda_0  B\wedge B\rangle}  e^{i(\l -\l_0 )\int_M \langle B\wedge B\rangle} ,
\end{gather*}
which means that our perturbation parameter is $\l -\l_0$ instead of~$\l$.

Let us consider the partial sums
\[
Z_P (\l,N) = \sum_{n=0}^P i^n Z_n (N) (\l -\l_0)^n  ,
\]
where $P =\sqrt{N}$. In this way we ensure that each perturbative order $n$ in $Z_P$  is much smaller than $N$ when $N$ is large. We can then use the estimates from the previous section, and
from~(\ref{zeva}) and~(\ref{zoda}) we obtain
\begin{gather*} {Z_P(\l,N)\over Z_0(\l_0)}  \approx \sum_{m=0}^{P/2} i^{2m} (\l-\l_0)^{2m} {N^m \over m!} z_2^m
+ \sum_{m=0}^{(P-3)/2} i^{2m+3} (\l -\l_0)^{2m+3} z_3 (z_2)^m \frac{ N^m}{m!} \nonumber\\
\phantom{{Z_P(\l,N)\over Z_0(\l_0)}}{}
\approx  \left[1 + i z_3 (\l-\l_0)^3 \right] \sum_{m=0}^\infty (-1)^m {g^m \over m!} z_2^m = \left[1 + i z_3 (\l-\l_0)^3 \right]e^{-gz_2}  ,
\end{gather*}
where
\begin{gather}
g=(\l -\l_0)^2 N  .\label{defg}
\end{gather}

Given that
\[
Z(\l,N) = \lim_{P\to\infty} Z_P (\l,N)
\]
 for $|\l -\l_0| < r$, where $r$ is the radius of convergence of
\begin{gather}
Z(\l,N)=\sum_{n=0}^\infty  i^n (\l-\l_0)^n Z_n(N)  ,\label{zln}
\end{gather}
then
\[
Z(\l,N) \approx Z_{\sqrt N} (\l,N )
\]
 for large $N$. Therefore
\begin{gather}{Z (\l,N) \over Z_0(\l_0)} \approx  \left[1 + i z_3 (\l-\l_0)^3 \right] \exp\left(- g z_2\right)   ,\label{zlnap}\end{gather}
for $\l \approx \l_0 $, where $g = (\l -\l_0)^2 N$. In the limit $N\to\infty$, $\l \to\l_0$ and $g$ constant
we obtain
\[
 {Z(\l,N) \over Z_0(\l_0)} \to \exp(-gz_2 )  .
 \]
We can rewrite this result as
\begin{gather} Z(M,g) = e^{-g z_2} Z_0(M) ,\label{dglz} \end{gather}
where $Z(M,g)$ is the perturbed partition function
in the dilute-gas limit. This value is triangulation independent and it depends on the renormalized coupling constant~$g$.

Note that (\ref{zlnap}) can be rewritten as
\begin{gather}
{Z^* (\l,g) \over Z_0(\l_0)} \approx  \left[1 + i z_3 (\l-\l_0)^3 \right] \exp\left(- g z_2\right)   ,\label{lgza}
\end{gather}
where we have changed the variable $N$ to variable  $g = N(\l -\l_0)^2$ and
\begin{gather}
Z^* (\l , g) = Z\left(\l, {g\over (\l -\l_0)^2}\right)  .\label{tz}
 \end{gather}
The approximation (\ref{lgza}) is valid for $\l \to\l_0$ and
\begin{gather*} {g \over (\l -\l_0)^2 } \to \infty  .
\end{gather*}
The result (\ref{dglz})  can be understood as the lowest order term in the asymptotic expansion (\ref{lgza}) where $g$ is a constant. However, one can have a more general situation where $g=f(\l -\l_0)$ such that
\begin{gather}{ f(\l -\l_0) \over (\l -\l_0)^2} \to \infty  ,\label{fc} \end{gather}
for $\l\to\l_0$.

In this case
\begin{gather} Z^*(\l,g) \approx e^{ -z_2 f(\l -\l_0)} \left[1 + iz_3 (\l -\l_0)^3 \right] Z_0 (\l_0)  ,\label{zlga}\end{gather}
which opens the possibility that
\begin{gather}  Z^* (\l,g) \approx \left[Z_0(k)\right]_ {k=\phi(\l)}  ,\label{eqf}\end{gather}
where $Z_0 (k)$ is the real number extension of the Crane--Yetter state sum and $k=\phi(\l)$ is the relation between $k$ and $\l$.

In the case of a quantum $SU(2)$ group at a root of unity
\begin{gather} Z_0 (M,k) = e^{-i\pi s(M) R(k)  }   , \label{cyz}\end{gather}
where $s(M)$ is the signature of $M$ and $R(k)= k(2k+1)/4(k+2)$,  see  {\cite{Ro}.} The value of $\l$ which corresponds to $k$ is conjectured to be $k = 4\pi/\lambda$, see \cite{smolin}, and this is an example of the function~$\phi$.

The relation $\l \propto 1/k$ could be checked by calculating the large-spin asymptotics of the quantum $15j$-symbol for the case of the root of unity, in analogy with the three-dimensional case, where by computing the asymptotics of the quantum $6j$-symbol one can f\/ind that $\l = 8\pi^2/k^2$,  see~\cite{MT}. The quantum $15j$-symbol asymptotics is not known, but for our purposes it is suf\/f\/icient to know that  $\lambda \to 0$ as $k \to\infty$.

{\sloppy Equations (\ref{zlga}), (\ref{eqf}) and (\ref{cyz}) imply
\begin{gather} f(\l -\l_0) \approx  i{\pi s(M) \over z_2} [h(\l) - h(\l_0)] + i{z_3 \over z_2} (\l-\l_0)^3  ,\label{fap}
\end{gather}
where $h(\l) = R(\phi(\l))$. The solution (\ref{fap}) is consistent with the condition~(\ref{fc}), since \mbox{$h'(\l_0) \ne 0$}. However, $f$ has to be a complex function, although the original def\/inition~(\ref{defg}) suggests a real function. This means that $\l$ has to take complex values in order for (\ref{eqf}) to hold, i.e.\ we need to perform an analytic continuation of the function in~(\ref{lgza}).

}

Also note that the def\/inition (\ref{zln}) and the fact that $Z_1 (N) = 0$ (Theorem~\ref{firstorder}) imply that $Z'(\l_0,N)=0$, but this does not imply that
\[
\lim_{\l\to\l_0}{d Z^* (\l ,f(\l -\l_0))\over d\l}= 0
\]
since $Z(\l,N)$ and $Z^* (\l,g)$ are dif\/ferent functions of $\l$ due to (\ref{tz}) and $g=f(\l -\l_0)$. Therefore the approximation (\ref{eqf}) is consistent. This also implies that we can def\/ine a triangulation independent $Z(\l,M)$ {as the real number extension of the function $Z_0 (k,M)$, see~(\ref{cyz})}. Therefore
\begin{gather}
Z(\l,M)  = Z_0 (k,M)|_{k=\f(\l)}= \tilde Z_0 (\l,M)  .\label{tiz}
\end{gather}

\section[Relation to $\langle F\wedge F\rangle $ theory]{Relation to $\boldsymbol{\langle F\wedge F\rangle}$ theory}\label{section6}

It is not dif\/f\/icult to see that the equations of motion for the action (\ref{abfp}) are equivalent to the equations of motion for the action
\begin{gather*} \tilde S = \int_M \langle F \wedge F \rangle  ,
\end{gather*}
because the $B$ f\/ield can be expressed algebraically as $B_I=-g_{IJ}F^J /(2\l)$.

At the path-integral level, this property is ref\/lected by the following consideration. One can formally perform a Gaussian integration over the $B$ f\/ield in the path integral (\ref{pibf}), which gives the following path integral
\begin{gather} Z = D(\l,M) \int {\cal D} A \exp\left( \frac{i}{4\l}\int_M  \langle F \wedge F \rangle \right) ,\label{ffpi}
\end{gather}
where $D(\l,M)$ is the factor coming from the determinant factor in the Gaussian integration formula.

More precisely, if we discretize $M$ by using a triangulation $T$ with $n$ triangles, then the path integral (\ref{pibf}) becomes a f\/inite-dimensional integral
\begin{gather}
\int \prod_{l,I} dA^I_l \int \prod_{f,I} dB^I_f \exp\bigg( i\sum_\D Tr\left(B_\D F_f\right) + \frac{i\l}{5} \sum_{\s} \sum_{\D',\D''\in \s}\langle B_{\D'} B_{\D''} \rangle\bigg) .\label{dpi}
\end{gather}

The integral over the $B$ variables in (\ref{dpi}) can be written as
\begin{gather}\int_{-\infty}^{+\infty}\cdots\int_{-\infty}^{+\infty}\prod_{ k,l}d B_{kl}   e^{i\l ({\cal B}  , Q  {\cal B}) + i({\cal B}  , {\cal F} )} ,\label{bint}\end{gather}
where $m=\dim A$, ${\cal B} = (B_{11},\dots,B_{mn})$ and ${\cal F}=(F_{11},\dots,F_{mn})$ are vectors in ${\bf R}^{mn}$, $(X,Y)$ is the usual scalar product in ${\bf R}^{mn}$ and $Q$ is an $mn\times mn$ matrix.

The integral (\ref{bint}) can be def\/ined as the analytic continuation $\l\to i\l$, ${\cal F}\to i{\cal F}$ of the formula
\begin{gather} \int_{-\infty}^{+\infty}\cdots\int_{-\infty}^{+\infty}\prod_{k,l}d B_{kl}   e^{-\l ({\cal B}  , Q  {\cal B}) + ({\cal F} , {\cal B}) } = \sqrt\frac{\pi^{mn}}{\l^{mn} \det  Q} e^{({\cal F} ,  Q^{-1}{\cal F})/4\l}  ,\label{gi}
\end{gather}
so that when $n\to\infty$ such that the triangulations become arbitrarily f\/ine, we can represent the limit as the path integral (\ref{ffpi}).

Since $\int_M  \langle F \wedge F\rangle $ is a topological invariant of $M$, which is the second characteristic class number $c_2(M)$, see~\cite{EGH}, we can write
\begin{gather*}
Z(M,\l) = E(M,\l) e^{i c_2(M)/\l}  ,
\end{gather*}
where
\[
 E(M,\l) = D(M,\l) \int {\cal D} A
 \]
and $D(M,\l)$ denotes the $(\l^{mn}\det Q)^{-1/2}$ factor from (\ref{gi}).
As we have shown in the previous section, $Z(\l,M) = \tilde Z_0 (M,\l)$, so that in the case of $SU(2)$
\begin{gather}
E(M,\l) = e^{-i c_2(M)/\l -i\pi s(M )h(\l)}   .\label{affcy}
\end{gather}

Therefore one can calculate the volume of the moduli space of connections on a principal bundle provided that the relation $k=\phi(\l)$ is known for the corresponding quantum group.

\section{Conclusions}
\label{section7}

The techniques developed for the 3d spin foam perturbation theory in \cite{MM2} can be extended to the 4d case, and hence the 4d partition function  has the same form as the corresponding 3d partition function in the dilute gas limit, see~(\ref{dglz}).
The constant $z_2$  depends only on the group~$G$ and an integer, and $z_2$ is related to the second-order perturbative contribution, see Section~\ref{five}. The constant $z_2$ appears because the constant $z_1$ vanishes for the same reason as in the 3d case, which is the vanishing {of} the tadpole spin network evaluation.

\looseness=1
The result (\ref{dglz}) implies that $Z(M,g)$ is not a new manifold invariant, but it is proportional to the Crane--Yetter invariant. Given that the renormalized coupling constant $g$ is an arbitrary number, a more usefull way of representing our result is the asymptotic formula~(\ref{zlga}). This formula allowed us to conclude that $Z(M,\l)$ can be identif\/ied as the Crane--Yetter partition function evaluated at $k=\phi(\l)$, see~(\ref{eqf}) and~(\ref{tiz}). The formula~(\ref{zlga}) also applies to the spin foam perturbation expansion in 3d, where $z_2$ and $z_3$ are given as the state sums of the corresponding 3d graphs, see~\cite{MM2}.  Therefore the formula (\ref{zlga}) is the justif\/ication for the conjecture made in~\cite{MM2}, where the triangulation independent $Z(\l,M)$ was identif\/ied with the Turaev--Viro partition function $Z_{TV} (M,k)$ for $k=4\pi^2 /\l^2$ in the $SU(2)$ case.

\looseness=1
The relation (\ref{tiz}) was useful for determining the volume of the moduli space of connections on the $G$-principal bundle for arbitrary values of $\l$, given that $Z(M,\l)$ is related to the path integral of the $ \langle F\wedge F\rangle$ theory, see~(\ref{affcy}). However, it still remains to be proved the conjecture that $k\propto 1/\l$ for $G=SU(2)$, while for the other groups the function $k=\phi (\l)$ is not known.

Note that the result~(\ref{dglz}) depends on the existence of a class of triangulations of $M$ which are arbitrarily f\/ine, but having a f\/inite degree of local complexity. As explained in Section~\ref{five} it is reasonable to assume that such a class exists, and can be constructed by considering the triangulations coming from the barycentric subdivisions of a f\/ixed cubulation of~$M$.

Our approach applies to Lie groups whose vector space of  intertwiners $\L\otimes\L\to A$ is one-dimensional for each irreducible representation~$\L$. This is true for  the $SU(2)$ and $SO(4)$ groups, but it is not true for the $SU(3)$ group. This can probably be f\/ixed by adding extra information to the chain-mail link with insertions at the 3-valent vertices.

Also note that we only considered the $g_{IJ} \propto \d_{IJ}$ case. This is suf\/f\/icient for simple Lie groups, but in the case of semi-simple groups one can have non-trivial $g_{IJ}$. Especially interesting is the $SO(4)$ case, where $g_{IJ} \propto \e_{abcd}$. In {the} general case one will have to work with spin networks which will have $L^{(\L)}$ and $g_{IJ}$ insertions, so that it would be very interesting to f\/ind out how to generalize the Chain--Mail formalism to this case.

One of the original motivations for developing a four-dimensional spin-foam perturbation theory was a possibility to obtain a nonperturbative def\/inition of the four-dimensional Euclidean quantum gravity theory, see~\cite{FK} and also~\cite{M1,M2}. The reason for this is that general relativity with a cosmological constant is equivalent to a perturbed $ BF$-theory
given by the action~(\ref{abfp}), where $G = SO(4,1)$ for a positive cosmological constant, while $G=SO(3,2)$ for a negative cosmological constant and $g_{IJ}=\e_{abcd}$ in both cases, see~\cite{M1,M2}. However, the $g_{IJ}$ in the gravity case is not a $G$-invariant tensor, since it is only invariant under a subgroup of $G$, which is the Lorentz group. Consequently this perturbed $BF$ theory is not topological.

In the Euclidean gravity case one has $G=SO(5)$, and the invariant subgroup is $SO(4)$ since $g_{IJ}=\e_{abcd}$. One can then formulate a spin foam perturbation theory along the lines of Section~\ref{foc}. However, the Chain--Mail techniques cannot be used, because $g_{IJ}$ is not a $G$-invariant tensor and therefore one lacks an ef\/f\/icient way of calculating the perturbative contributions. In order to make  further progress, a generalization of  the Chain--Mail
calculus has to be found in order to accommodate the case when~$g_{IJ}$ is
invariant only under a subgroup of~$G$.

\subsection*{Acknowledgments}
This work was partially supported FCT (Portugal) under the projects  $\rm PTDC/MAT\!/\!099880/2008$,
 PTDC/MAT/098770/2008,
 PTDC/MAT/101503/2008.
This work was also partially supported by CMA/FCT/UNL, through the project
PEst-OE/MAT/UI0297/2011.

\pdfbookmark[1]{References}{ref}
\LastPageEnding


\begin{thebibliography}{99}

\footnotesize\itemsep=1.5pt

\bibitem{Ba}
Baez J.,
An introduction to spin foam models of quantum gravity and BF theory,
in Geometry and Quantum Physics (Schladming, 1999),
\href{http://dx.doi.org/10.1007/3-540-46552-9_2}{{\it Lecture Notes in Phys.}}, Vol.~543, Springer, Berlin, 2000, 25--93,
\href{http://arxiv.org/abs/gr-qc/9905087}{gr-qc/9905087}.

\bibitem{Ba2}
Baez J.,
Spin foam perturbation theory, in Diagrammatic Morphisms and Applications (San Francisco, CA, 2000),
{\it Contemp. Math.}, Vol.~318, Amer. Math. Soc., Providence, RI, 2003,  9--21,
\href{http://arxiv.org/abs/gr-qc/9910050}{gr-qc/9910050}.


\bibitem{BGM}
Barrett J.W.,  Faria Martins J.,  Garc{\'\i}a-Islas J.M.,
Observables in the Turaev--Viro and Crane--Yetter models,
\href{http://dx.doi.org/10.1063/1.2759440}{{\it J.~Math. Phys.}} {\bf  48}  (2007),   093508, 18~pages,
\href{http://arxiv.org/abs/math.QA/0411281}{math.QA/0411281}.

\bibitem{B}
Broda B.,
Surgical invariants of four-manifolds,
\href{http://arxiv.org/abs/hep-th/9302092}{hep-th/9302092}.

\bibitem{CT}
Cooper D.,  Thurston W.,
 Triangulating $3$-manifolds using $5$ vertex link types,
\href{http://dx.doi.org/10.1016/0040-9383(88)90004-3}{{\it Topology}} {\bf  27}  (1988),    23--25.


\bibitem{CY}
Crane L., Yetter D.A.,
A categorical construction of $4$D topological quantum f\/ield theories,
 in Quantum Topology, \href{http://dx.doi.org/10.1142/9789812796387_0005}{{\it Ser. Knots Everything}}, Vol.~3, World Sci. Publ., River Edge, NJ, 1993,  120--130,
\href{http://arxiv.org/abs/hep-th/9301062}{hep-th/9301062}.

\bibitem{EGH}
Eguchi T.,  Gilkey P.B.,  Hanson A.J.,
Gravitation, gauge theories and dif\/ferential geometry,
\href{http://dx.doi.org/10.1016/0370-1573(80)90130-1}{{\it Phys. Rep.}} {\bf  66}  (1980), 213--393.



\bibitem{MM}
Faria Martins  J., Mikovi\'{c} A.,
Invariants of spin networks embedded in three-manifolds,
\href{http://dx.doi.org/10.1007/s00220-008-0422-8}{{\it Comm. Math. Phys.}} {\bf 279} (2008), 381--399,
\href{http://arxiv.org/abs/gr-qc/0612137}{gr-qc/0612137}.

\bibitem{MM2}
Faria Martins  J., Mikovi\'{c} A.,
Spin foam perturbation theory for three-dimensional quantum gravity,
\href{http://dx.doi.org/10.1007/s00220-009-0776-6}{{\it Comm. Math. Phys.}} {\bf  288}  (2009),   745--772,
\href{http://arxiv.org/abs/0804.2811}{arXiv:0804.2811}.

\bibitem{FK}
Freidel L., Krasnov  K.,
Spin foam models and the classical action principle,
{\it  Adv. Theor. Math. Phys.} {\bf 2} (1999), 1183--1247,
\href{http://arxiv.org/abs/hep-th/9807092}{hep-th/9807092}.

\bibitem{FS}
Freidel L., Starodubtsev A.,
Quantum gravity in terms of topological observables,
\href{http://arxiv.org/abs/hep-th/0501191}{hep-th/0501191}.

\bibitem{GS}
Gompf R.E., Stipsicz  A.I.,
 $4$-manifolds and Kirby calculus,
 {\it Graduate Studies in Mathematics}, Vol.~20, American Mathematical Society, Providence, RI, 1999.

\bibitem{KL}
Kauf\/fman L.H., Lins S.L.,
Temperley--Lieb recoupling theory and invariants of 3-manifolds,
{\it Annals of Mathematics Studies}, Vol.~134, Princeton University Press,  Princeton, NJ, 1994.

\bibitem{K}
Kirby  R.C.,
 The topology of $4$-manifolds,
 {\it Lecture Notes in Mathematics}, Vol.~1374, Springer-Verlag, Berlin, 1989.

\bibitem{L}
Lickorish W.B.R.,
The skein method for three-manifold invariants,
\href{http://dx.doi.org/10.1142/S0218216593000118}{{\it J.~Knot Theory Ramifications}} {\bf  2} (1993), 171--194.


\bibitem{Ma1}
Mackaay M.,
Spherical $2$-categories and $4$-manifold invariants,
\href{http://dx.doi.org/10.1006/aima.1998.1798}{{\it Adv. Math.}} {\bf  143} (1999),  288--348,
\href{http://arxiv.org/abs/math.QA/9805030}{math.QA/9805030}.

\bibitem {Ma2}
Mackaay M.,
Finite groups, spherical 2-categories, and 4-manifold invariants,
\href{http://dx.doi.org/10.1006/aima.1999.1909}{{\it Adv. Math.}} {\bf  153}  (2000),  353--390,
\href{http://arxiv.org/abs/math.QA/9903003}{math.QA/9903003}.

\bibitem{M3}
Mikovi\'c A.,
Spin foam models of Yang--Mills theory coupled to gravity,
\href{http://dx.doi.org/10.1088/0264-9381/20/1/317}{{\it Classical Quantum Gravity}} {\bf 20}  (2003),   239--246,
\href{http://arxiv.org/abs/gr-qc/0210051}{gr-qc/0210051}.

\bibitem{M1}
Mikovi\'c A.,
Quantum gravity as a deformed topological quantum f\/ield theory,
\href{http://dx.doi.org/10.1088/1742-6596/33/1/029}{{\it J.~Phys. Conf. Ser.}} {\bf 33} (2006), 266--270,
\href{http://arxiv.org/abs/gr-qc/0511077}{gr-qc/0511077}.

\bibitem{M2}
Mikovi\'c A.,
Quantum gravity as a broken symmetry phase of a BF theory,
\href{http://dx.doi.org/10.3842/SIGMA.2006.086}{{\it SIGMA}} {\bf 2} (2006), 086, 5~pages,
\href{http://arxiv.org/abs/hep-th/0610194}{hep-th/0610194}.

\bibitem{MT}
Mizoguchi S., Tada T.,
Three-dimensional gravity from the Turaev--Viro invariant,
\href{http://dx.doi.org/10.1103/PhysRevLett.68.1795}{{\it Phys. Rev. Lett.}} {\bf 68} (1992), 1795--1798,
\href{http://arxiv.org/abs/hep-th/9110057}{hep-th/9110057}.

\bibitem{RT}
Reshetikhin  N., Turaev V.G.,
Invariants of $3$-manifolds via link polynomials and quantum groups,
\href{http://dx.doi.org/10.1007/BF01239527}{{\it  Invent. Math.}} {\bf  103}  (1991),   547--597.

\bibitem{Ro}
Roberts J.,
Skein theory and Turaev--Viro invariants,
\href{http://dx.doi.org/10.1016/0040-9383(94)00053-0}{{\it Topology}} {\bf  34}  (1995),    771--787.

\bibitem{RS}
Rourke C.P., Sanderson B.J.,
Introduction to piecewise-linear topology, Reprint, {\it Springer Study Edition}, Springer-Verlag, Berlin~-- New York, 1982.

\bibitem{smolin}
Smolin L.,
Linking topological quantum f\/ield theory and nonperturbative quantum gravity,
\href{http://dx.doi.org/10.1063/1.531251}{{\it J.~Math. Phys.}} {\bf 36} (1995), 6417--6455,
\href{http://arxiv.org/abs/gr-qc/9505028}{gr-qc/9505028}.

\bibitem{T}
Turaev V.G.,
Quantum invariants of knots and 3-manifolds,
{\it de Gruyter Studies in Mathematics}, Vol.~18, Walter de Gruyter \& Co., Berlin, 1994.

\bibitem{W}
Witten E.,
Quantum f\/ield theory and the Jones polynomial,
\href{http://dx.doi.org/10.1007/BF01217730}{{\it Comm. Math. Phys.}} {\bf  121}  (1989),  351--399.

\end{thebibliography}
\end{document}